\documentclass[preprint,showpacs,aps,amsmath,amssymb]{revtex4}

\usepackage{bm}
\usepackage{graphicx}
\begin{document}


\title{On the low-temperature lattice thermal transport in nanowires}

\author{Alexander V. Zhukov}
\author{Shilong Yang}
\author{Jianshu Cao} 

 \affiliation{Department of Chemistry, Massachusetts Institute of Technology,
              Cambridge, MA 02139, USA}

\date{\today}

\begin{abstract}

We propose a theory of low temperature thermal transport in
nano-wires in the regime where a competition between phonon and
flexural modes governs the relaxation processes. Starting with the
standard kinetic equations for two different types of
quasiparticles we derive a general expression for the coefficient
of thermal conductivity. The underlying physics of thermal
conductance is completely determined by the corresponding
relaxation times, which can be calculated directly for any
dispersion of quasiparticles depending on the size of a system. We
show that if the considered relaxation mechanism is dominant, then
at small wire diameters the temperature dependence of thermal
conductivity experiences a crossover from $T^{1/2}$ to
$T^3$-dependence. Quantitative analysis shows reasonable agreement
with resent experimental results.

\end{abstract}


\maketitle

Low-dimensional materials have attracted considerable attention in
recent years, particularly, in view of their potential
applications in electronic devices \cite{ratner,light}. Many
theoretical and experimental studies of nanowires and nanotubes
are centered on the properties of electronic transport. However it
is realized now that the thermal properties of nanomaterials are
also important for applications \cite{leitner1,leitner2,hui2}. It
is of special interest to increase thermal conductance in the
micro and nanodevices \cite{cui,pon,ratner}. In this paper study
thermal transport in nonmetallic systems, in which heat is
transported by thermal excitations only. In addition to practical
importance of such studies, thermal transport in nanowires is
interesting from fundamental point of view. Recent theoretical
\cite{1} and experimental \cite{2} findings proved the existence
of the quantum of thermal conductance in ballistic regime, which
is similar to the quantum of electronic conductance. The state of
experimental and theoretical understanding of thermal transport in
nanoscale systems is comprehensively discussed in the review
\cite{3}. Recently D. Li et al \cite{4} reported an accurate
measurement of lattice thermal conductivity in silicon nanowires
for a wide range of temperatures and wire diameters. They
demonstrated the significant influence of the system size not only
on the magnitude of the thermal conductivity coefficient, but also
on its temperature dependence. It is well known that for large
enough diameters of the wire and diffusive phonon - boundary
scattering, thermal conductivity coefficient at low temperatures
is proportional to $T^3$. But for small values of the wire
diameters experiment \cite{4} shows clear crossover from cubic to
near linear dependence on the temperature. In the present paper we
consider one particular relaxation mechanism which can explain the
observed crossover.

Recently Mingo \cite{5} carried out an accurate numerical study of
thermal conductance of silicon nanowires to explain the decrease
of the thermal conductivity coefficient with wire diameter
observed in the experiment. He assumed that all the effects can be
explained by the reconstruction of the phonon dispersion, where
realistic phonon modes obtained from MD simulations were applied
to general expression of the thermal conductivity coefficient. His
numerical analysis shows excellent quantitative agreement with the
experiment \cite{4} for large enough diameters at high
temperatures. As the system size becomes smaller, the approach
fails to describe a sharp decrease of thermal conductance as well
as qualitative change of its temperature dependence. This is
likely because the Matheissen's rule has been used for evaluation
of the phonon lifetime, which has rather restricted range of
applicability (see e.g. \cite{ad3} and references therein).

As it was noted in Ref. \cite{glavin}, decrease of the temperature
increases the characteristic phonon wavelength and reduces the
scattering probability at the boundary surface. This leads to a
modification of the phonon spectrum. For ideal wires it is
represented by a set of branches with energies proportional to 1D
momentum directed along the wire. So the standard theory of
thermal conductance in dielectrics and semiconductors has to be
modified to account for low dimensionality effects as well as
phonon spectrum modification at low temperatures. Thus, to
understand thoroughly the physical processes occurring inside the
nanowires with decreasing sizes, we need an analytical theory to
account for different mechanisms explicitly, such as dispersion
reconstruction and restricted geometry.

To approach the problem we consider low enough temperatures, where
the quasiparticle states of "acoustic" branches are thermally
populated ($\epsilon \longrightarrow 0$ when $p \longrightarrow
0$). The corresponding acoustic branches have the following
dispersion relations \cite{6,glavin}:

\begin{equation}
\epsilon_1 = u_1 p_1, \qquad \epsilon_2 = u_2 a p_2^2, \label{1}
\end{equation}
where $\epsilon_i$ stands for the energy of a quasiparticle, $p_i$
is the corresponding momentum, $a$ is the wire diameter, $u_1$ and
$u_2$ are the characteristic velocities. The first expression in
Eq. (\ref{1}) is the phonon dispersion and the second expression
is the dispersion of flexural mode. The nature of flexural modes
comes from the fact they are analogous to bending modes of
classical elasticity theory, or the antisymmetric Lamb waves of a
free plate \cite{auld}. Appearance of such a mode is just a direct
consequence of restricted geometry, and under some conditions it
can be considered as the only size effect in thermal transport
properties. Strictly speaking there are two modes for each type of
dispersion, but we do not account for them separately since their
contributions are qualitatively the same. Consequently, we have to
solve the kinetic problem for two-component gas of quasiparticles.
It is well known \cite{ad1,ad2,ad3,ad4,7} that using simple
Callaway formula to estimate the thermal conductivity coefficient
in two-component systems sometimes leads to serious confusions.
Simple summation of the relaxation rates, as is done in the
majority of theoretical works, is questionable under many physical
conditions. The dependence of the kinetic coefficients on
different relaxation times is much more complicated in reality.
Accurate method for calculation of the diffusion coefficient in
two-component gas of quasiparticles was proposed in Ref. \cite{7}.
Here we extend this formalism to thermal conductance problem.

To start, we consider a system of two types of quasiparticles.
Their kinetics is described by equations for corresponding
distribution functions $f_i$:
\begin{equation}
v_i\frac{\partial f_i}{\partial z}=\sum_{j=1}^2 C_{ij} (f_i,f_j) +
C_{i3}(f_i), \qquad i=1,2. \label{2}
\end{equation}
where $C_{ij} (f_i,f_j)$ is the collision integral of thermal
excitations, and $ C_{i3}(f_i)$ is the collision integral
describing the scattering processes between quasiparticles and
scatterers. In general the latter includes all the processes
leading to the non-conservation of the total energy. $v_i =
{\partial \epsilon_i}/{\partial p_i}$ is the group velocity of the
corresponding thermal excitation. The main purpose of our theory
is to obtain analytic expressions of thermal conductance, which
are applicable to quasiparticles with arbitrary dispersion
relations. In other words, the explicit dispersion relations in
Eq. (\ref{1}) are needed only at the last stage when calculating
corresponding relaxation times and thermodynamic quantities. As
usual, we seek a perturbative solution of the system (2) in the
form
\begin{equation}\label{3}
f_i = f_i^{(0)} + \delta f_i,
\end{equation}
where $f_i^{(0)}$ is the local equilibrium Bose-function and
$\delta f_i \ll f_i^{(0)}$ represents small deviation from the
equilibrium. The perturbation term can be conveniently chosen to
be $\delta f_i = -g_i {\partial f_i^{(0)}}/{\partial\epsilon_i}$
with $g_i$ the new target functions. After the standard
linearization procedure, Eq. (\ref{2}) can be written in the
following matrix form:
\begin{equation} \label{5} \vert \phi_\kappa\rangle \frac{1}{T}
\frac{\partial T}{\partial z} ={\hat{\cal{C}}} \vert g \rangle,
\end{equation}
where
$$
\vert \phi_\kappa\rangle = \left\vert
\begin{array}{l}
\epsilon_1 v_1 \\
\epsilon_2 v_2
\end{array}\right\rangle, \quad
\vert g \rangle =\left\vert\begin{array}{l}
g_1 \\
g_2
\end{array} \right\rangle.
$$
The 2D collision matrix ${\hat{\cal{C}}}$ can be decomposed into a
sum of three terms, corresponding to different relaxation
mechanisms: $ {\hat{\cal{C}}} = {\hat{\cal{J}}} + {\hat{\cal{S}}}
+ {\hat{\cal{U}}}, $ where ${\hat{\cal{J}}}$, with matrix elements
${\cal{J}}_{ij}=C_{ik}\delta_{ij}+C_{ij}(1-\delta_{ij})$ ($k \neq
i$), describes the relaxation due to interaction between
quasiparticles of different types; ${\hat{\cal{S}}}$
(${\cal{S}}_{ij}=C_{ii}\delta_{ij}$) describes collisions between
identical quasiparticles, and ${\hat{\cal{U}}}$
(${\cal{U}}_{ij}=C_{i3}\delta_{ij}$) describes all the other
relaxation mechanisms, which do not conserve total energy of the
quasiparticle system. ${\hat{\cal{U}}}$ includes scattering on
defects, boundaries, umklapp processes etc. \cite{hui} Here
$C_{ij}$ represent {\it linearized} collision operators.

Let us define the scalar product of two-dimensional bra- and
ket-vectors as follows \cite{7}:

\begin{equation}\label{11}
\langle \phi \vert \chi \rangle =\sum_{k=1,2}(\phi_k \vert \chi_k)
=  - \sum_{k=1,2}\int \phi_k^* \chi_k \frac{\partial
f_k^{(0)}}{\partial\epsilon_k} d\Gamma_k,
\end{equation}
where $(\phi_k \vert$ and $\vert \chi_k)$ are the correspondent
one-component vectors, $d\Gamma$ is the element of phase volume.
Under this condition, the collision operator ${\hat{\cal{C}}}$
becomes hermitian. System (\ref{5}) is the system of nonuniform
linear integral equations. According to the general theory of
integral equations the target solution $\vert g \rangle$ must be
orthogonal to the solution of corresponding uniform equations
${\hat{\cal{C}}}\vert \phi_{uni} \rangle$. It is therefore
convenient to write the formal solution of (\ref{5}) so that the
orthogonality condition $\langle g \vert \phi_{uni} \rangle$ is
imposed explicitly in the solution. For this purpose we define the
projection operator ${\hat{\cal{P}}}_n$ onto the subspace
orthogonal to the vector $\vert\phi_{uni}\rangle$,
${\hat{\cal{P}}}_n = 1 - {\hat{\cal{P}}}_c, \quad
{\hat{\cal{P}}}_c = \vert \phi_{uni}\rangle \langle \phi_{uni}
\vert.$ As a result, the formal solution of the system (\ref{5})
can be written in the form

\begin{equation}\label{13}
\vert g \rangle = {\hat{\cal{P}}}_n \left( {\hat{\cal{C}}}^{-1}
\right) {\hat{\cal{P}}}_n \vert \phi_\kappa\rangle \frac{1}{T}
\frac{\partial T}{\partial z}.
\end{equation}

The heat flux density due to the thermal excitations of different
types is given by the expression $Q = \sum_{k=1,2}\int \epsilon_k
v_k f_k d \Gamma_k.$ Using relation (\ref{3}) and definition of
scalar product (\ref{11}), $Q$ can be rewritten as $ Q = \langle
\phi_{\kappa}\vert g \rangle.$ On the other hand, the effective
thermal conductivity coefficient is defined by the relation $ Q =
- \kappa_{eff}{\partial T}/{\partial z}. $ Comparing the above two
expressions for $Q$, and using the formal solution (\ref{13}) we
obtain

\begin{equation}\label{17}
\kappa_{eff} = -\frac{1}{T}\langle\phi_{\kappa}\vert
{\hat{\cal{C}}}^{-1}\vert \phi_{\kappa} \rangle .
\end{equation}
To derive an exact and analytical expression for thermal
conductivity coefficient (\ref{17}) it is necessary to introduce a
complete set of orthonormal two-dimensional vectors $\vert \psi_n
\rangle$ ($n=1,2,3,...$) belonging to the infinite-dimensional
Hilbert space with scalar product (\ref{11}). In principle, the
particular choice of the basis is not essential, but for the
convenience of calculations it is useful to specify at least four
of them. It's convenient to chose the first of them to correspond
to the total momentum of quasiparticles and the second one to be
orthogonal, but still linear in momentum \cite{7}:
\begin{equation}\label{18}
\vert \psi_1 \rangle = \frac{1}{\sqrt{\rho}} \left\vert \begin{array}{l} p_1 \\
p_2 \end{array} \right\rangle, \quad \vert \psi_2 \rangle =
\frac{1}{\sqrt{\rho\rho_1\rho_2}} \left\vert \begin{array}{l} \rho_2 p_1 \\
-\rho_1 p_2
\end{array} \right\rangle,
\end{equation}
where $\rho_i = (p_i\vert p_i)$ is the normal density of the $i$th
component, $\rho = \rho_1 +\rho_2$. The third and the fourth
vectors correspond to the energy flux:

\begin{equation}\label{20}
\vert \psi_3 \rangle = \frac{1}{{\cal N}_{\kappa 1}} \left\vert \begin{array}{l} \psi_{\kappa 1} \\
0 \end{array} \right\rangle, \quad \vert \psi_4 \rangle =
\frac{1}{{\cal N}_{\kappa 2}} \left\vert \begin{array}{l} 0 \\
\psi_{\kappa 2} \end{array} \right\rangle,
\end{equation}
where
\begin{equation}\label{21}
\psi_{\kappa {j}} = \frac{1}{\sqrt{T}} \left( \epsilon_j v_j -
\frac{S_jT}{\rho_j}p_j\right),
\end{equation}
and ${\cal N}_{\kappa {j}}= \sqrt{(\psi_{\kappa {j}} \vert
\psi_{\kappa {j}})}$ is the corresponding normalization
coefficient. Partial entropy of quasiparticle subsystem $S_j$ in
eq. (\ref{21}) is given by the relation
\begin{equation}\label{23}
S_j = \frac{1}{T} (\epsilon_j v_j \vert p_j).
\end{equation}

Formally, the kinetic problem of two-component quasiparticles
system can be solved in the above basis set. The inversion of the
operator matrix ${\hat{\cal{C}}}$ in Eq. (\ref{17}) is similar to
the procedure described in Ref. \cite{7}. The final result
contains infinite dimensional non-diagonal matrices. To obtain
closed form expressions we must use some approximations, correct
$\tau$-approximation \cite{ad4} or Kihara approximation
\cite{7,kihara,mason}. In some physical situations we are able to
obtain closed analytical expressions. It is rigorously proved in
\cite{7} that in case of quasi-equilibrium within each subsystem
of quasiparticles, the corresponding transport coefficient can be
obtained in close analytical form. This is a reliable
approximation when the low temperature relaxation is mainly
governed by the defect scattering processes. The approximation
formally implies that all the matrix elements of matrix $\hat{\cal
S}$ in eq. (\ref{17}) tend to infinity. The thermal conductivity
coefficient in this case can be obtained in the form:
$\kappa_{eff} = \kappa_F + \kappa_D$. Here we separate the flux
part of thermal conductivity coefficient $\kappa_F =
\tau_F{S^2T}/{\rho}$ with $S=S_1+S_2$, which approaches infinity
when the quasiparticles do not interact with scatterers, and the
diffusive part $\kappa_D = \tau_D\left( {S_1T}/{\rho_1} -
{S_2T}/{\rho_2} \right)^2 {\rho_1 \rho_2}/{T \rho}$. The
corresponding relaxation times are given by

\begin{equation}\label{27}
\tau_D = \left\{\frac{\rho_1}{\rho}\tau_{23}^{-1} +
\frac{\rho_2}{\rho}\tau_{13}^{-1} + \tau_{12}^{-1} +
\tau_{21}^{-1}\right\}^{-1},
\end{equation}
and

\begin{equation}
\tau_F = \tau_D \left( \frac{S_1}{S}\tau_{23}^{-1} +
\frac{S_2}{S}\tau_{13}^{-1} + \tau_{12}^{-1} + \tau_{21}^{-1}
\right)^2
(\tau_{13}^{-1}\tau_{23}^{-1}+\tau_{12}^{-1}\tau_{23}^{-1}+\tau_{21}^{-1}\tau_{13}^{-1})^{-1}
\label{28}.
\end{equation}

Relaxation times contained in formulas (\ref{27}) - (\ref{28}) are
defined by
\begin{equation}\label{29}
\tau_{kj}^{-1} = \frac{1}{\rho_k}(p_j\vert C_{kj} \vert p_j ).
\end{equation}
We emphasize that they are not actual scattering times, which are
momentum dependent, but are relaxation times associated with
corresponding scattering mechanisms. Once we obtain the particular
scattering rate $\nu_{kj}(p_k)$ from standard scattering theory,
we can replace true collision operator $C_{k}$ with
$\nu_{kj}(p_k)$, so that the corresponding relaxation time can be
calculated by
\begin{equation}
\tau_{kj}^{-1}= \rho_k^{-1}\int p_k^2 \nu_{kj}(p_k)\frac{\partial
f_k^{(0)}}{\partial \epsilon_k}d\Gamma_k \label{30}
\end{equation}

As can be seen from the derived formulas, the coefficient of
thermal conductivity contains different relaxation times in rather
non-trivial combination. If drop out one component (say set
$S_2=0$, $\rho_2=0$) we recover the usual result $
\kappa_{F}^{(1)} = \tau_{13}{S_1^2T}/{\rho_1}$. For phonons with
linear dispersion $\epsilon = vp$, $\kappa_F$ reduces to the
well-known result $ \kappa_F^{ph} = C_{ph}v^2\tau_{1 3}/3 $, where
$C_{ph}=3S_{ph}$ is the heat capacity of phonon gas.

The main advantage of our approach is its universality. In fact
till now we have not restricted ourselves to any particular
dimensionality of the system or any quasiparticles dispersion. All
the necessary information is contained in the corresponding
relaxation times and thermodynamic quantities. This formalism
allows us to analyze contributions from different relaxation
mechanisms to the total thermal conductivity coefficient. Given
the dispersion relations of quasiparticles, we can easily
calculate all the quantities contained in (\ref{27}) - (\ref{28}).

With Eqs. (\ref{27}) - (\ref{28}) we are able to address the
competition between relaxation processes of the flexural and
phonon modes. Glavin \cite{glavin} noted that such a competition
can be essential at extremely low temperatures if the dominant
relaxation mechanism is elastic scattering on defects, where he
argued that thermal conductivity coefficient would scale as
$T^{1/2}$. Our approach allows us to study this competition
comprehensively. Particularly, we predict the strong dependence of
the temperature scaling exponent on the wire diameter. The
standard Fermi golden rule approach \cite{glavin} gives the
momentum dependent scattering rates for different modes $ \nu_{13}
= W_{13}\frac{p^{1/2}}{u_1^3a^{3/2}}$, $\quad \nu_{23} =
W_{23}\frac{1}{p u_2^3a^3},$ where $W_{kj}$ are the corresponding
scattering amplitudes, which depend on the physical properties of
particular material. Using Eq. (\ref{30}) it is easy to show, that
the corresponding relaxation times scale as
\begin{equation}\label{35}
\tau_{13}^{-1} \propto a^{-3/2}T^{1/2}, \quad \tau_{23}^{-1}
\propto a^{-5/2}T^{- 1/2}.
\end{equation}

Different temperature dependence of relaxation times lead to a
strong competition between two physically different mechanisms of
thermal conductivity, flux and diffusive. The dominance of one
over the other strongly depends on the wire diameter at a given
temperature. To make some specific conclusions let us summarize
the approximations done and specify the range of validity of the
proposed theory. We consider a situation when thermal excitations
are multiply scattered elastically while being transferred through
the wire, so that other scattering mechanisms are strongly
suppressed by interaction with defects. Only for this case we were
able to drop relaxation within each subsystem of identical
quasiparticles to obtain closed expressions (\ref{27}),
(\ref{28}). The influence of boundary is accounted in the
dispersion of flexural mode and in the dimensionality of the
system. The range of temperature is supposed to satisfy the
relation $T<\Delta\epsilon$, where $\Delta\epsilon\sim 1/a$ is the
characteristic value of the frequency gap between the adjacent
phonon branches. For larger temperature we cannot use the acoustic
modes (\ref{1}) only, but need to account higher branches.

In Figure 1 we compare our theoretical results with the
experimental data from \cite{4}. We have chosen the unknown
parameters $W_{13}=1.2\cdot 10^{-44}\rm{m^5s^{-4}}$ and
$W_{23}=0.9\cdot 10^{-44}\rm{m^5s^{-4}}$ to fit data for $a=22$nm.
Deviations from the experimental data for large diameters and
temperatures show the restriction of applicability of our initial
approximations. They arise from the Debye approximation and
simplified dispersion expression. Additionally, when the diameter
of the wire increases, the mechanism we consider becomes less
dominant. To be more precise we need to include higher excitation
branches as well as another relaxation mechanisms. However our
approach allows to understand the physics of the processes in the
region under consideration. Clearly, observed crossover is the
result of competition between $\kappa_F \propto T^{1/2}$ and
$\kappa_D \propto T^3$. For smaller diameters, $\kappa_F$ is
strongly dominant in the wide range of temperatures as shown on
Figure 2 for wire diameter $a=2$nm. Figure 3 demonstrates a
complete crossover from $T^{1/2}$ to $T^3$ dependence for the
nanowire of $a=30$nm. It can be seen that the $T$ dependence
between $20$K and $40$K is nearly linear, which was observed in
the experiment \cite{4}. It should be noted that $T^3$ dependence
of $\kappa_D$ cannot be interpreted in simple analogy with the
bulk case. It comes not from a specific heat directly, but from
different sources including competition of the relaxation times in
Eqs. (\ref{27}), (\ref{28}).

In summary, we derived the general analytical expressions
(\ref{27}) - (\ref{28}), to explicitly calculate the contributions
of different scattering mechanisms to the total relaxation of the
system. The simple expressions clarify the essential effects
leading to the observed behavior of thermal conductivity
coefficient. It is clear that the particular dispersion laws (and
their reconstruction) affect scattering rates and thermodynamic
quantities. Restricted geometry and low dimensionality lead to
additional scattering mechanisms. Note the information about
dimensionality is naturally included in the particular form of
phase space element $d \Gamma_i$. Such a formalism helps to
distinguish effects from different scattering mechanisms. When
applied to the regime, where phonon modes compete with flexural
ones, our theory agrees favorably with available experimental
data. Furthermore, we showed that the thermal conductivity
coefficient changes from approximately $T^{1/2}$-dependence to
$T^3$-dependence with increasing temperature. In view of our
theoretical results it is useful to investigate smaller diameters
or lower temperatures with fixed diameters in experiment to better
reveal the crossover from $T^3$ to $T^{1/2}$ dependence.

We would like to thank Prof. Peidong Yang and Dr. Deyu Li for
providing us the experimental data. This research is supported by
the NSF Career Award (Grant No.~Che-0093210). J.~C. is a recipient
of the Camille Dreyfus Teacher-Scholar Award.

\newpage
\begin{figure}
  \includegraphics[width=6cm]{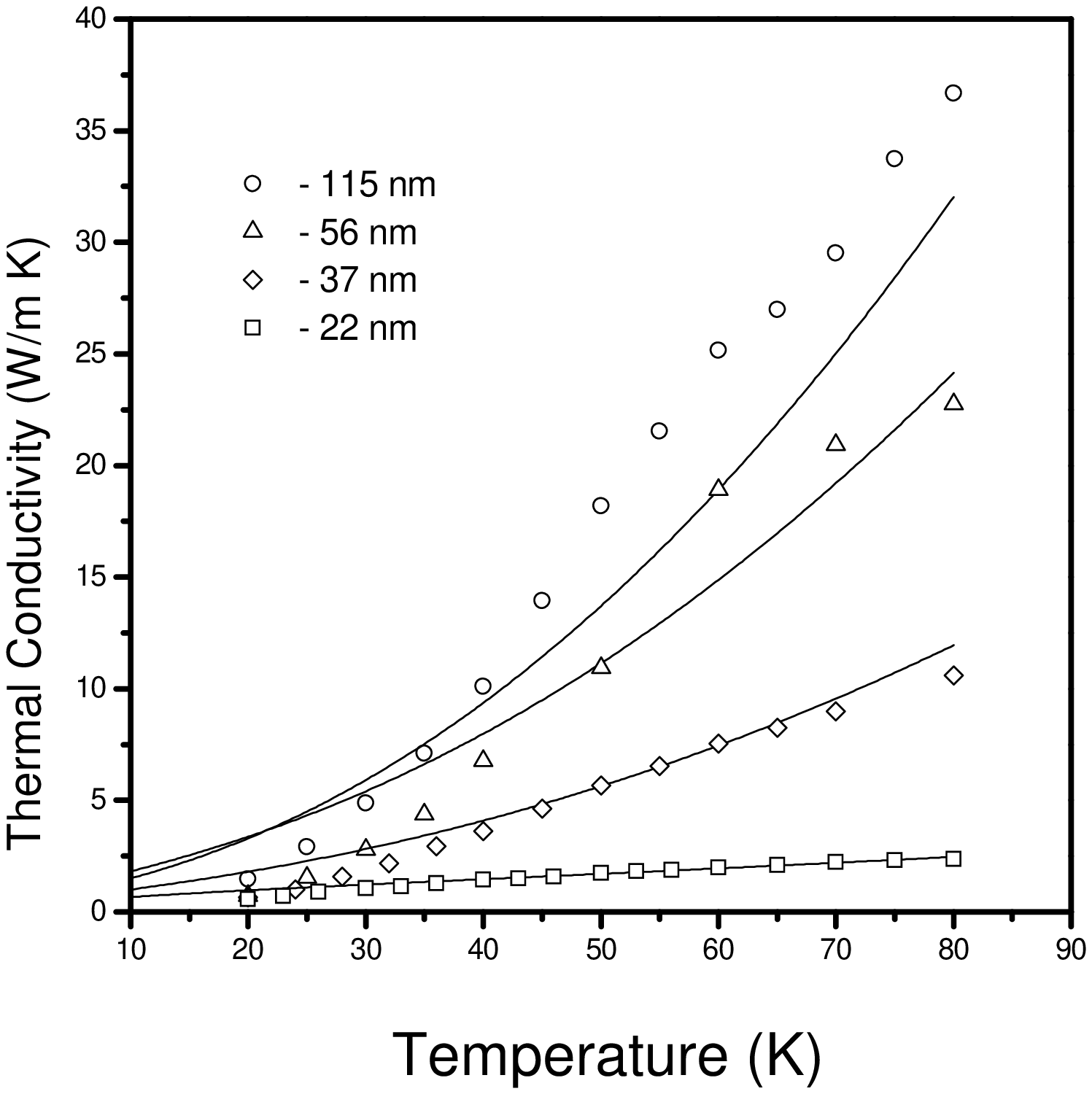}\\
  \caption{Thermal conductivity coefficient calculated from Eqs. (\ref{27})
  and (\ref{28}) for different values of nanowire diameter. Experimental
data are from Ref. \cite{4}.}\label{fig1}
\end{figure}

\begin{figure}
  \includegraphics[width=8cm]{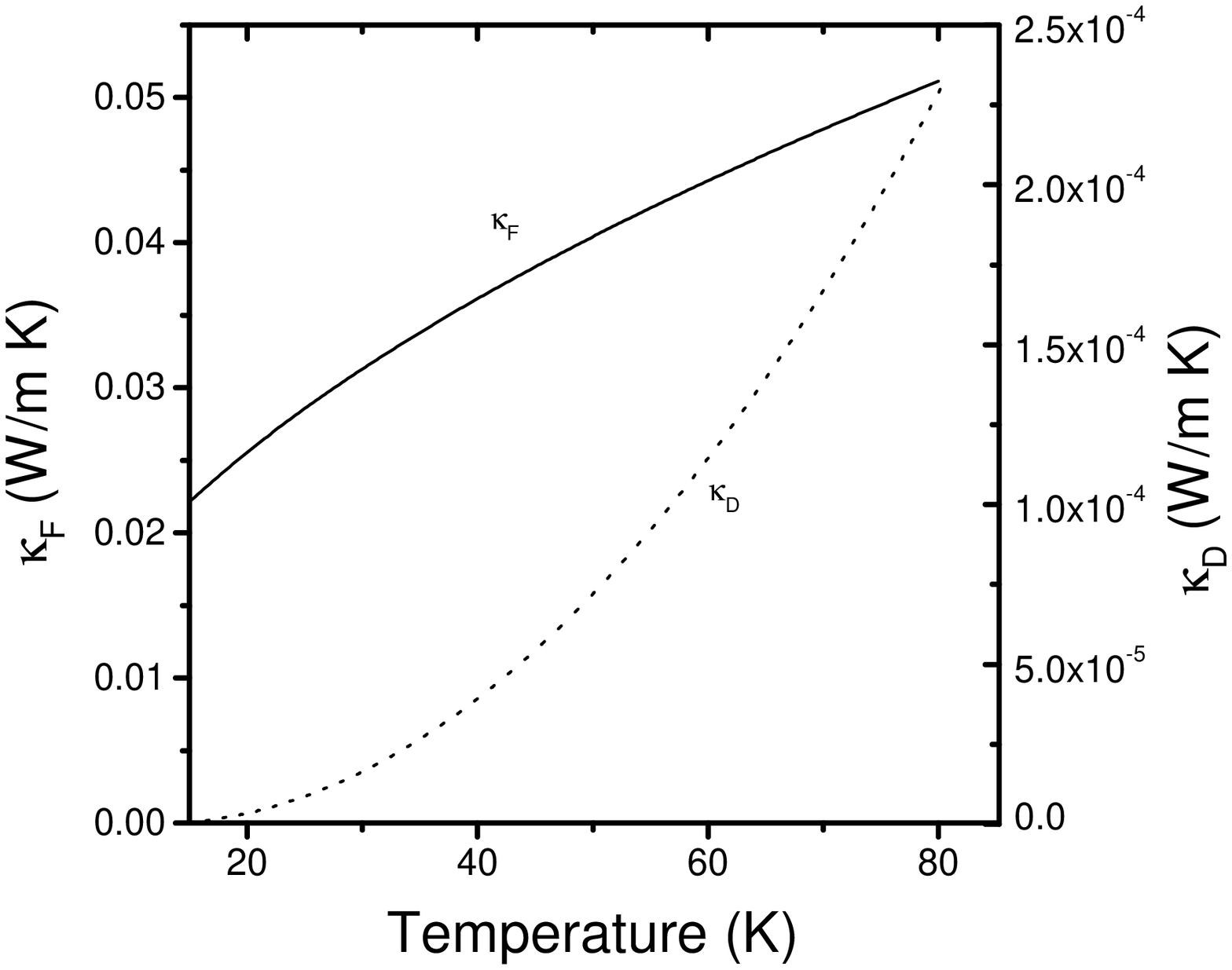}\\
  \caption{Comparative contribution from flux and diffusive parts of
thermal conductivity for a 2nm wire.}\label{fig2}
\end{figure}

\begin{figure}
  \includegraphics[width=6cm]{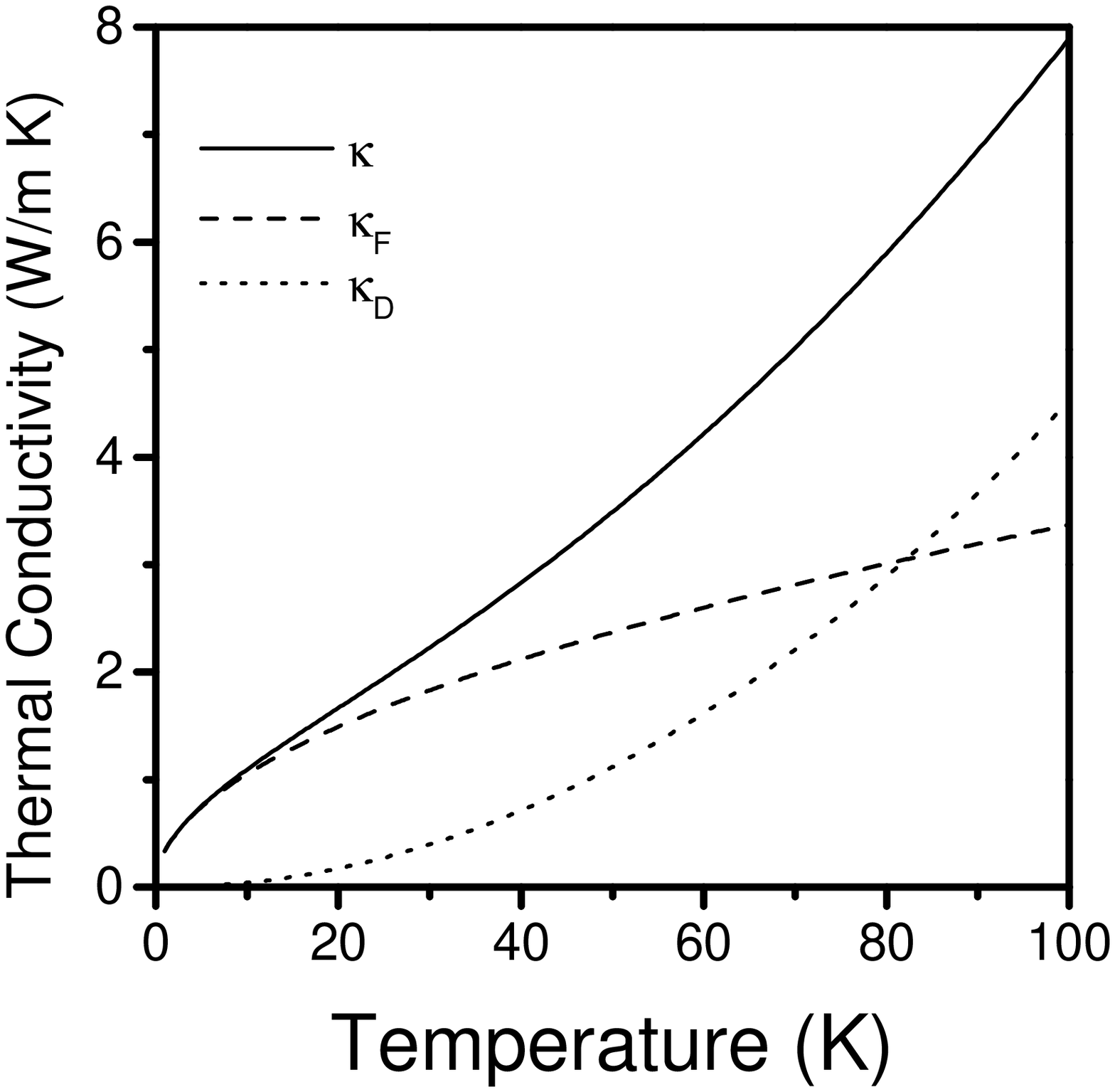}\\
  \caption{Crossover from $T^{1/2}$ to $T^3$ dependence of $\kappa_{eff}$
for a 30nm wire.}\label{fig3}
\end{figure}

\end{document}